\documentclass[12pt,preprint]{aastex}

\shorttitle{Absence of BL AGNs in CGs}
\shortauthors{Mart\'{\i}nez et al.}

\begin{document}

\title{Deficiency of Broad Line AGNs in Compact Groups of Galaxies}

\author{M. A. Mart\'{\i}nez\altaffilmark{1} and A. Del Olmo\altaffilmark{1}}
\affil{Instituto de Astrof\'{\i}sica de Andaluc\'{\i}a, CSIC, Apdo. 3004,
18080 Granada, Spain}
\email{geli@iaa.es, chony@iaa.es}
\author{R. Coziol\altaffilmark{2} }
\affil{Departamento de Astronom\'{\i}a, Universidad de Guanajuato,
Apdo. 144, 36000 Guanajuato, Mexico}
\email{rcoziol@astro.ugto.mx}
\and
\author{P. Focardi\altaffilmark{3}}
\affil{Dipartimento di Astronomia, Universit{\`a} di Bologna, via
Ranzani 1, 40127 Bologna, Italy}
\email{paola.focardi@unibo.it}

\begin{abstract}
Based on a new survey of AGN activity in Compact Groups of Galaxies,
we report a remarkable deficiency of Broad Line AGNs as compared to
Narrow Line AGNs. The cause of such deficiency could be related to
the average low luminosity of AGNs in CGs: $10^{39}$ erg s$^{-1}$.
This result may imply lower accretion rates in CG AGNs, making Broad
Line Regions (BLR) undetectable, or may indicate a genuine absence
of BLRs. Both phenomena are consistent with gas stripping through
tidal interaction and dry mergers.
\end{abstract}


\section{Introduction}
From the optical spectra of AGNs, one can generally distinguish two
main types: those that show broad emission lines (BLAGNs) and those
that show only narrow emission lines (NLAGNs). With an absolute
magnitude M$_V\geq -23$, the local BLAGNs are called Seyfert~1 (Sy1),
while the NLAGNs are called Seyfert~2 (Sy2). In the literature, one
can also find other types of Seyfert galaxies: Sy1.2, Sy1.5, Sy1.8,
and Sy1.9, all of them being some sort of Sy1 and consequently
BLAGNs. NLAGNs may also come into the form of Low Ionization Nuclear
Emission-line Regions (LINERs).

Phenomenologically, it is unclear why AGNs come in different types.
Based on spectral variation, the Narrow Line Region (NLR) is thought
to be located farther out from the central Black Hole accretion disk
than the Broad Line Region (BLR), and to be spatially much more
extended. Within the unification model \citep{ant93}, which assumes
all AGNs to be intrinsically the same, the BLR in NLAGNs is hidden
behind an optically thick torus of gas and dust. Consistent with
this model, many observations of NLAGNs have revealed hidden BLRs
through polarized spectroscopy \citep{ant02}. However, not all
NLAGNs observed with this technique have shown such structures
\citep{tra01,tra03,lao03,shu07}, suggesting that in some NLAGNs the
BLR might simply be absent. This last finding is consistent with
alternative models in which the accretion rate and, consequently,
AGN luminosity plays a direct role in determining the presence of
the BLR \citep{nic00,nic03,eli06}.

One possible way how to solve this dilemma is to explore the
connection of AGNs with their environment. According to the
unification model, for example, one does not expect to find any
differences in the number of AGNs in different environments.
Unfortunately, such studies are usually controversial. While some
authors found Sy2 to inhabit richer environments than Sy1
\citep{rob98}, others claimed the opposite \citep[and references
therein]{sch01}. Recently, \citet{kou06} have found no differences
in the environment of Sy1 and Sy2 over large scales ($\lesssim$  1
Mpc), but a higher fraction of Sy2 with close companion than Sy1 in
systems with spatial scales smaller than 100 kpc (using H$_0$=70 km
s$^{-1}$ Mpc$^{-1}$). These results agree with \citet{sor06} who
found two times more Sy2 than Sy1 in local ($\lesssim$ 100 kpc)
environment.

To explore further the possible conection between AGN activity and
environment we have undertaken a new survey to determine the nature
and frequency of nuclear activity in two different samples of
Compact Groups of galaxies (CGs): the Hickson Compact Groups
\citep[HCGs]{hic82} and a sample of CGs from the Updated Zwicky
Catalog \citep[UZC-CG]{foc02}. Previous studies on CGs revealed a
high percentage of low luminosity AGNs in these structures, but very
few luminous ones \citep{coz98,coz00,mar06,mar07}. Having in hand a
statistically significant sample of CGs with complete information on
the nuclear activity of their members, allows us to better quantify
the frequency of BLAGNs in these systems.

\section{Data and Results}
\label{sec2}

Among the HCGs, we have selected the groups with redshifts
$z\leq0.045$, having a surface brightness $\mu_G\leq24.4$. These
criteria provided us with a statistically complete sample of 283
galaxies in 65 groups. We have obtained medium resolution
spectroscopy for 238 of these galaxies. The spectra of 71 galaxies
come from previous observations made by \citet{coz98,coz00,coz04}.
The remaining 167 galaxies were observed by our group using four
different telescopes: the 2.5m NOT\footnote{ALFOSC is owned by the
IAA and operated at the Nordic Optical Telescope (NOT), under
agreement between IAA and the NBIfAFG of the Astronomical
Observatory of Copenhagen.} in ``El Roque de los Muchachos'' (RM,
Spain), the 2.2m in Calar Alto (CAHA\footnote{The Centro
Astron{\'o}mico Hispano Alem\'an is operated jointly by the
Max-Planck Institut fur Astronomie and the IAA-CSIC.}, Spain), the
2.12m in San Pedro M{\'a}rtir (SPM, Mexico) and the 1.5m telescope
in Sierra Nevada Observatory (OSN, Spain).

For all the galaxies, broad components search and activity
classification were done only after template subtraction, to correct
for absorption features produced by the underlying stellar
population. Detailed of the template subtraction method used can be
found in \citet{coz98,coz04}. Preliminary results have already been
published in \citet{mar07}. Complete description of observations,
reduction and analysis method will be published elsewhere.

For the UZC-CG sample, we have collected spectra from three
spectroscopic archives: the Sloan Digital Sky Survey (SDSS-DR4), the
Z-Machine and the FAST Spectrograph Archives. We have found spectra
for all the galaxy members of 215 groups (720 galaxies): 210 spectra
are from the SDSS, 187 from FAST and 323 from Z-Machine
\citep{mar06}. Because
the Z-Machine spectra have too low S/N ratios to measure broad
components, we restrict our analysis to the 397 spectra found in the
SDSS and FAST databases. Spectra from the SDSS survey
were template subtracted using \citet[H05]{hao05} eigenspectra and
their PCA method. No correction was applied to the galaxies in the
FAST sample, due to the non availability of suitable spectra to be used
as templates.

\begin{figure}
\epsscale{1.1} 
\plotone{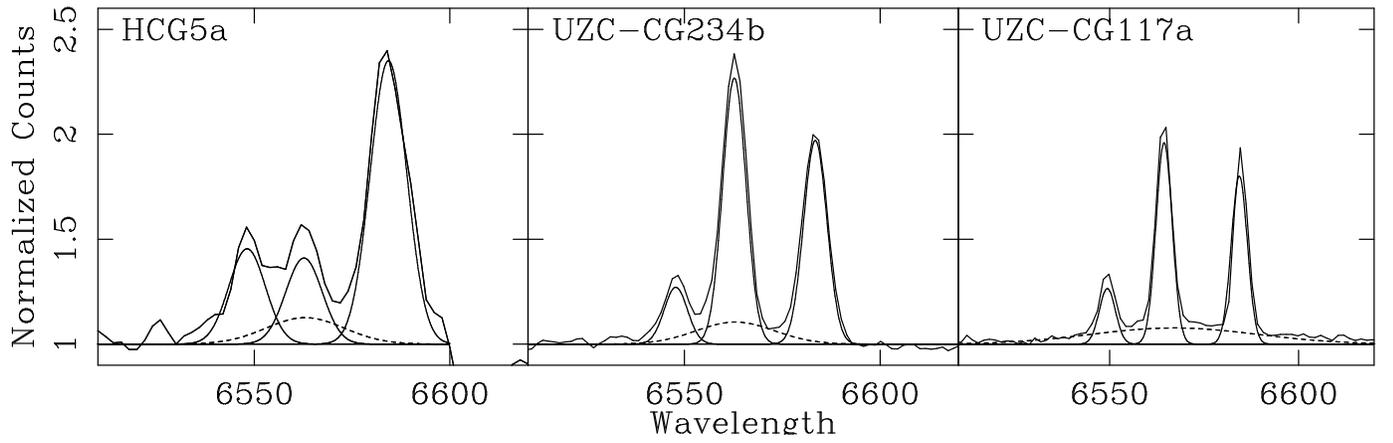} 
\caption{Examples of BLAGNs in
our sample with multi-components Gaussian decompositions centered on
H$\alpha$.  \label{fig1}}
\end{figure}

Emission lines were found in 153 of the 238 galaxies in the HCG
sample (64\%) and 274 of the 397 galaxies (69\%) in the UZC-CG
sample. The identification of broad components was done by fitting a
multi-components Gaussian on the emission lines, using the IRAF
task NGAUSSFIT. The FWHM of [SII] (or [OIII] when the [SII] lines
were too faint or noisy) have been used to model the narrow
components of the H$_{\alpha}$ and [NII] lines. When a broad fourth
component was necessary, it was centered on H$_{\alpha}$. A $\chi^2$
criterion, as described in H05, was applied to choose the fourth
component parameters, establishing in this way its presence and
characteristics. Examples of fitting plots for three BLAGNs are
shown in Fig~\ref{fig1}.

Following \citet{ost89} we classified BLAGNs galaxies having FWHM
$\ge 500$ km s$^{-1}$. Our analysis revealed only 1 BLAGN in the HCG
sample and 8 in the UZC-CG sample. For each of these galaxies we give,
in Table ~\ref{tbl1} the FWHM of the broad
component and activity classification according to
\citet{ost89}: a Sy1.9 shows a broad component only in H$_{\alpha}$,
while a Sy1.8 shows also a weak broad component in H$_{\beta}$. None
of our BLAGNs are classified as Sy1, Sy1.2 or Sy1.5. Based on our
analysis, BLAGNs represent only 1\% (1/153) of emission line
galaxies in the HCG sample and 3\% (8/274) in the UZC-CG one.

To classify NLAGNs we used the diagnostic diagram
based on the four most intense emission lines: H$_{\beta}$,
[OIII]5007\AA\ , H$_{\alpha}$, and [NII]6583\AA\ and criteria
similar to those employed by \citet{kew06}. Galaxies located above
the theoretical maximum sequence for star formation are classified
as AGNs. We also distinguished between Sy2 and LINERs using the
classical upper limit log([OIII]5007\AA/H$_{\beta}$) $<$ 0.4 for
LINER. Both CG samples are rich in galaxies having only
[NII]6583\AA\ and H$_{\alpha}$, we classified these cases as Low
Luminosity AGNs (LLAGNs) when log([NII]/H$_{\alpha}$)$>-0.1$
\citep{coz98,sta06}.

A summary of the nuclear classification for the AGN galaxies in our
two samples are presented in Table~\ref{tbl2}. For each sample
(column 1) we give the number of Sy2, LINERs and LLAGNs (columns 2
to 4), which, put together, constitute the total NLAGN populations.
In column 5 we report the fractions of BLAGNs over NLAGNs and in
column 6 the fraction of Sy1 over Sy2, considering all BLAGNs as Sy1
like.

\subsection{Lower ratio of BLAGNs to NLAGNs in CGs than in the field}

The fraction of BLAGNs over NLAGNs in our two CG samples is
extremely low: 1\% for the HCGs and 6\% for the UZC-CGs. Also
noticeably low are the ratios of Sy1 to Sy2: 4\% in the HCG and 19\%
in the UZC-CG. To realize how low these ratios are one has to
compare with what is usually found in other surveys.

The mean H$_{\alpha}$ luminosity of both NLAGNs and BLAGNs in our two
samples is about $10^{39}$ erg s$^{-1}$, which is typical of the faint
end of the luminosity function of AGNs. This value is comparable with
the mean H$_{\alpha}$ luminosity of AGNs observed in the local
universe by \citet[HFS97]{ho97a}. Except for some galaxies in Virgo,
all the galaxies in the HFS97 sample are located in low density
environments (either loose groups or isolated). In Table~\ref{tbl2} we
compare their results (395 galaxies, excluding the galaxies in Virgo)
with ours. In the HFS97 sample, the BLAGNs were classified as such by
\citet{ho97b}, based on the detection of a broad H$_{\alpha}$
component. To be consistent with our definition, all these galaxies
were classified as Sy1. Also for comparison sake, the narrow emission
lines galaxies in the HFS97 sample were reclassified using the
criteria described in sect.~\ref{sec2}.

The fraction BLAGN/NLAGN in the HFS97 sample is 22\% and the ratio
Sy1/Sy2 is 61\%. There is consequently a clear deficiency of BLAGNs
in CGs. This also appears as an extremely large difference in the
number of Sy1 as compared to Sy2 galaxies. This phenomenon is quite
intriguing considering that there is no deficit of AGNs as a whole
in CGs: 46\% AGNs in the HCG, 51\% in the UZC-CG compared to 44\% in
the HFS97 sample.

Comparable ratios (Sy1/Sy2 $\backsim$ 60\%) were obtained by
\citet[SRR06]{sor06} in the field, with a slight increase in ``loose
groups'' (Sy1/Sy2 $\backsim$ 69\%). In the nearby (z$<$0.33) sample
of SDSS AGN galaxies covering four orders of luminosity and similar
environments as SRR06, H05 determined a ratio BLAGN/NLAGN of 43\%
and a ratio Sy1/Sy2 of 54\%. Assuming BLAGNs are slightly favored at
higher luminosity these high ratios are comparable to those found by
HFS97.

\section{Discussion}

\subsection{Quantifying biases and detection limits}

Our results suggest there is an important deficit of BLAGNs in our
two CG samples as compared to similar surveys in the field. This
result confirm the tendency first encountered by
\citet{coz98,coz00}. To verify that the lack of BLAGNs in CGs can not
be induced by differences in observation, reduction or
analysis methods we have investigated thoroughly these
possibilities.

Comparison of the UZC-CG sample with the survey made by H05 is safe,
because our SDSS data derive from the same telescope, reduction and
analysis methods (including template subtraction) as theirs. The ratio
BLAGN/NLAGN is 8\% in our sample compared to 43\% for the sample of
H05, which is already a huge difference.

A possible effect due to difference in spectral resolution can also be
excluded. \citet{ho97b} have used high resolution (2.5 \AA) spectra,
but made tests with two others low resolution set-ups (5\AA\ and
10\AA) obtaining similar results. These are comparable to our own
observations: CAHA and OSN (4\AA), NOT and SPM (8\AA), SDSS (3.5\AA)
and FAST (6\AA). Both H05 and SRR06 have 3.5\AA.

The S/N continuum levels of the different surveys are also
comparable. On average the S/N of AGNs in our spectra is 60 with a
maximum of the order of 120. This is comparable to H05 and SRR06
spectra (they both used SDSS). HFS97 did not published their values.
However their BLAGNs rates are comparable to those of H05 and SRR06,
suggesting this is not an issue.

There is no evidence either for a higher galaxy contamination (the
amount of galaxy falling into the aperture) in our samples. Taking
into account the slit aperture and distances of the host galaxies in
each sample we find medians of 1 kpc and 1.3 kpc for the HCG and
UZC-CG, respectively. Although the median for HFS97 is lower (0.5
kpc) than for H05 (7 kpc) the results are similar. Obviously,
template subtraction (like we also did) alleviates the
differences. We may note also that no relation is observed, in any
of these surveys (including ours), between the frequency of BLAGNs
and the redshift of the galaxies where they are found, which means
that nearby galaxies are not more likely BLAGNs than remote ones.

To test if our low number of Sy1 could be due to a difference in
morphologies \citep{sch01}, we have divided our two samples and the
HFS97 one in three morphology classes: E for early-type galaxies
(E-S0), Se for early-type spirals (S0a-Sbc), and Sl for late-type
spirals (Sc and later). For homogeneity sake, all the morphologies
have been taken from the Hyperleda database \citep{pat03}. In
Table~\ref{tbl3} we give for each morphology class the fraction of
galaxy and the ratios BLAGN/NLAGN and Sy1/Sy2. There are no BLAGNs
in late-type spirals in any sample. In the HFS97 sample, the ratio
of BLAGN/NLAGN is marginally higher in the E class while the ratio
Sy1/Sy2 is significantly higher, which indicates a definitive
increase in BLAGNs in early-type galaxies. In the two CG samples we
almost see an inverse trend: the ratios of BLAGN/NLAGN and Sy1/Sy2
are both larger in the Se class than in the
E one. Moreover there is a definite rise
in the number of early-type galaxies in CGs. Following the HFS97
trend, this should have produced more BLAGNs in CGs instead of less.
This eliminates a difference in morphologies as a possible
explanation.

We also reject the hypothesis of lower sensitivity. Comparing the
median luminosity in H$\alpha$ of the different types of galaxies in
our samples with those in the HFS97 sample, lower sensitivity would
have translated into higher values in our samples. This is not
observed. In the HFS97 sample the median H$\alpha$ luminosity of the
NLAGNs is log(L$_{H\alpha} =38.72 $ ergs s$^{-1}$).  Our values are
comparable: 38.69 for the HCG and 38.79 for the UZC-CG.

Finally we have determined the detection limits in our samples as in
\citet{ho97b}. Different simulations were performed adding to each
set-up spectra a grid of synthetic spectra with broad gaussian
components of various widths and amplitudes centered on H$\alpha$.  We
then applied our template and extraction analysis to deduce the
following limits. For the CAHA and OSN spectra, broad components
equivalent to 15\% or higher of the total blended flux in
H$\alpha+$[NII] were recovered. Using medians of AGN blended flux and
redshfit this transforms into a detection limit in H$\alpha$ broad
luminosity of $3.5\times10^{38}$ ergs s$^{-1}$. We find slightly
higher fraction (20\%) for the NOT and SPM spectra, equivalent to a
detection limit in luminosity of $4.0\times10^{38}$ ergs
s$^{-1}$. Only three BLAGNs in the HFS97 sample have a luminosity
lower than these limits. Obviously, the lack of BLAGNs encountered in
our samples cannot be explain by a higher detection limits in our
samples.

There seem to be no obvious observational biases or differences in
reduction and analysis methods capable of reproducing the lack of
BLAGNs in CGs as compared to lower density environments. It is
consequently reasonable to conclude that this phenomenon must be
related to the environment typical of CGs.

\subsection{The disappearance of BLRs in CGs}

In the unification model for AGNs, a torus of matter is assumed to
be responsible for hiding the BLR from the observer. To be
consistent with our analysis, this mechanism should be much more
efficient in CGs. However, this assumption goes against the evidence
of tidal stripping: in CGs the infall of gas in the disk seem to be
stopped, generally diminishing the amount of star formation
\citep{coz98,coz00,ver01,ros07,dur08}. At the same time, the number
of early-type galaxies in CGs is observed to increase. Therefore, a
possible reason why no BLRs appear in AGNs in CGs may be because any
amount of gas that has reached the center was consumed into stars,
building larger bulges \citep{car99}. Alternatively, the bulges of
galaxies in CGs may have grown without gas, through dry mergers
\citep{coz07}.

The fact that the average luminosity of the AGNs in CGs is low is
another argument in favor of the dissolution hypothesis for the BLR.
According to recent results obtained by reverberation mapping, the
size of the BLR in AGNs is correlated to the optical luminosity at
5100\AA\ \citep{kas05}. It is consequently possible to imagine the
size of the BLR shrinking almost to zero at some low threshold
luminosity \citep{eli06}.  The luminosity at 5100\AA\ in our samples
range from log($\lambda$L$_{\lambda}$(5100\AA)) 40.7 to 43.1 (in
units of erg s$^{-1}$); comparing with data of \citet{pet04} we are
in the lower luminosity part of their distribution, where few
objects with broad lines have been observed. We also are at the
lower limit where no hidden BLRs have been found
\citep{shu07,bia07}. Using the relation log(L$_{bol}$/L$_{Edd}$),
most of our galaxies are below -1.37, which suggests that broad
features may simply not exist in these LLAGNs.

According to \citet{nic00} and \citet{nic03} low accretion rates
rather than smaller mass black holes are responsible in explaining
the absence of BLRs in Low Luminosity AGNs which is fully compatible
with our observations.

\section{Conclusion}

Based on the above statistics, we confirm that there is a remarkable
deficiency of BLAGNs as compared to NLAGNs in CGs. This result
suggests that BLRs in AGN CGs are directly affected by tidal or
group interaction effects, which make them shrink below detection or
completely disappear.

In CGs environment, galaxies are undergoing morphological
transformations and the main mechanisms for such transformations are
tidal interactions and mergers \citep{coz07}. Our analysis suggests
that the combined effects of these two mechanisms also result in an
important decrease in the amount of gas that can reach the nucleus
to form a BLR in AGNs.

\acknowledgments We are grateful to Lei Hao for making available her
eigenspectra and to Jaime Perea for his PCA software and helpful
discussion. MAM acknowledges Ministerio de Educacion y Ciencia for
financial support grant FPU AP2003-4064. MAM and AdO are partially
supported by spanish research projects AYA2006-1325 and TIC114. RC was
partially supported by the CONACyT, under grant
No. 47282. P.F. acknowledges financial contribution from MIUR and from
the contract ASI-INAF I/023/05/0. We thank the referee for
constructive comments. We also thank the TAC of the Observatorio
Astron\'omico Nacional at San Pedro M\'artir for time allocations. We
thank the SDSS collaboration for providing the extraordinary database
and processing tools that made part of this work possible. The SDSS
Web Site is http://www.sdss.org/. We acknowledge also the usage of the
Hyperleda database (http://leda.univ-lyon1.fr).

\clearpage

\begin{deluxetable}{lcccc}
\tablewidth{0pt}
\tablecaption{BLAGNs identification\label{tbl1}}
\tablehead{
Name         &    Source      & Type &   FWHM(H$_{\alpha}$) & FWHM(H$_{\beta}$)\\
             &                &      &   (km/s)       & (km/s)
}
\startdata
HCG 5a       & CAHA     & 1.9  & 1056   &\nodata \\
UZC-CG 84c   & SDSS     & 1.9  & 2727   &\nodata\\
UZC-CG 89b   & SDSS     & 1.9  & 2159   &\nodata\\
UZC-CG 109b  & SDSS     & 1.8  & 1902   & 1499 \\
UZC-CG 117a  & SDSS     & 1.9  & 2351   &\nodata\\
UZC-CG 132b  & FAST     & 1.9  & 3055   &\nodata\\
UZC-CG 139b  & SDSS     & 1.9  & 1941   &\nodata \\
UZC-CG 232c  & SDSS     & 1.8  & 2258   & 1689  \\
UZC-CG 234b  & FAST     & 1.9  & 1328   &\nodata\\
\enddata
\end{deluxetable}

\clearpage

\begin{deluxetable}{lcccccc}
\tablecaption{Nuclear classification for the AGN galaxies\label{tbl2}}
\tablewidth{0pt}
\tablehead{
Sample &  \multicolumn{3}{c}{NLAGN}   & BLAGN & $\frac{\rm{BLAGN}}{\rm{NLAGN}}$ & $\frac{\rm{Sy1}}{\rm{Sy2}}$ \\ \cline{2-4}
          & Sy2 & LINER & LLAGN     &   &  \%    & \% }
\startdata
HCG     & 28   & 23      &  19     &    1 &  1    &  4\\
UZC-CG  & 43   & 11      &  79     &    8 &  6    & 19 \\
HFS97   & 46   & 80      &\nodata  &   28 & 22    & 61 \\
H05     & 2424 & 650     &\nodata  & 1317 & 43    & 54 \\
SRR06   & 1104 &\nodata  &\nodata  &  725 & \nodata & 66  \\
\enddata
\end{deluxetable}

\clearpage

\begin{deluxetable}{lcccccccc}
\tablecaption{Activity types and morphological distribution\label{tbl3}}
\tablewidth{0pt} \tablehead{
Sample &\multicolumn{3}{c}{E-S0}&& \multicolumn{3}{c}{S0a-Sbc} & Sc-Irr\\ \cline{2-4}\cline{6-8}
       & f$_{E}$ &$\frac{\rm{BLAGN}}{\rm{NLAGN}}$ & $\frac{\rm{Sy1}}{\rm{Sy2}}$& & f$_{Se}$& $\frac{\rm{BLAGN}}{\rm{NLAGN}}$ & $\frac{\rm{Sy1}}{\rm{Sy2}}$ & f$_{Sl}$}
\startdata
HCG   &54\% & \nodata   &  \nodata  & &32\% &   3\%      &  10\% & 14\%\\
UZC   &34\% &  2\%      &  8\%      & &55\% &   9\%      &  25\% & 11\%\\
HFS97 &25\% & 25\%      & 86\%      & &39\% &  19\%      &  56\% & 36\% \\
\enddata
\end{deluxetable}

\clearpage

\begin{thebibliography}{}

\bibitem[Antonucci(1993)]{ant93} Antonucci, R.  1993, ARA\&A, 31, 473
\bibitem[Antonucci(2002)]{ant02} Antonucci, R.  2002, Astrophysical Spectropolarimetry,
(Cambridge, CUP), 151 (astro-ph/010348)
\bibitem[Bian \& Gu(2007)]{bia07} Bian, W., \& Gu, Q. 2007, \apj, 657, 159
\bibitem[Caon et al.(1994)]{cao94} Caon, N., Capaccioli, M., D'Onofrio, M., \& Longo, G. 1994, \aap, 286, 39
\bibitem[Coziol et al.(1998)]{coz98} Coziol, R., Ribeiro, A. L. B., Capelato, H. V., \& de~Carvalho, R. R. 1998, \apj, 493, 563
\bibitem[Coziol et al.(2000)]{coz00} Coziol, R., Iovino, A., \& de~Carvalho, R. R. 2000, \aj, 120, 47
\bibitem[Coziol et al.(2004)]{coz04} Coziol, R., Brinks, E., \& Bravo-Alfaro, H. 2004, \aj, 128, 68
\bibitem[Coziol \& Plauchu-Frayn(2007)]{coz07} Coziol, R., \& Plauchu-Frayn 2007, \aj, 133, 2630
\bibitem[de~Carvalho \& Coziol(1999)]{car99} de~Carvalho, R. R., \& Coziol, R. 1999, \aj, 117, 1657
\bibitem[de la Rosa et al.(2007)]{ros07} de la Rosa, I. G., de Carvalho, R. R., Vazdekis, A., \& Barbuy, B 2007, \aj, 133, 330
\bibitem[de Robertis et al.(1998)]{rob98} de Robertis, M. M., Yee, H. K. C., \& Hayhoe, K. 1998, \apj, 496, 93
\bibitem[Durbala et al.(2008)]{dur08} Durbala, A., et al. 2008, \aj, 135, 130
\bibitem[Elitzur \& Shlosman(2006)]{eli06} Elitzur, M., \& Shlosman, I. 2006, \apj, 648, L101
\bibitem[Focardi \& Kelm(2002)]{foc02} Focardi, P., \& Kelm, B.\ 2002, \aap, 391, 35
\bibitem[Hao et al.(2005)]{hao05} Hao, L., et al.\ 2005, \aj, 129, 1783 (H05)
\bibitem[Hickson(1982)]{hic82} Hickson, P.\ 1982, \apj, 255, 382
\bibitem[Hickson et al.(1988)]{hic88} Hickson, P., Kindl, E., \& Huchra J. P. 1988, \apj, 331, 64
\bibitem[Ho et al.(1997a)]{ho97a} Ho, L.~C., Filippenko, A.~V., \& Sargent, W.~L.~W.\ 1997a, \apjs, 112, 315 (HFS97)
\bibitem[Ho et al.(1997b)]{ho97b} Ho, L.~C., Filippenko, A.~V., Sargent, W.~L.~W., \& Peng, C.~Y. \ 1997b, \apjs, 112, 414
\bibitem[Kaspi et al.(2005)]{kas05} Kaspi, S., Maoz, D., Netzer, H., Peterson, B. M., Vestergaard, M., \& Jannuzi, B. T. 2005, \apj, 629, 61
\bibitem[Kewley et al.(2006)]{kew06} Kewley, L.~J., Groves, B., Kauffmann, G., \& Heckman, T.\ 2006, \mnras, 372, 961
\bibitem[Koulouridis et al.(2006)]{kou06} Koulouridis, E., Plionis, M., Chavushyan, V., Dultzin-Hacyan,D., Krongold, Y., \& Goudis, C.  2006, \apj, 639, 37
\bibitem[Laor(2003)]{lao03} Laor, A.\ 2003, \apj, 590, 86
\bibitem[Mart\'{\i}nez et al.(2006)]{mar06} Mart\'{\i}nez, M. A., del Olmo, A., Focardi, P., \& Perea J. 2006, VII Scientific Meeting of SEA (astro-ph/0611099)
\bibitem[Mart\'{\i}nez et al.(2007)]{mar07} Mart\'{\i}nez, M. A., del Olmo, A., Perea, J., \& Coziol, R. 2007, ESO Astro. Symp., Springer-Verlag, 163
\bibitem[Mendes de Oliveira \& Hickson(1994)]{men94} Mendes de Oliveira, C., \& Hickson, P. 1994, \apj, 427, 684
\bibitem[Nicastro(2000)]{nic00} Nicastro, F.\ 2000, \apjl, 530, L65
\bibitem[Nicastro et al.(2003)]{nic03} Nicastro, F., Martocchia, A., \& Matt, G.\ 2003, \apjl, 589, L13
\bibitem[Osterbrock(1989)]{ost89} Osterbrock, D.~E.\ 1989, in Astrophysics of Gaseous Nebulae and Active Galactic Nuclei, University Science Books
\bibitem[Paturel et al.(2003)]{pat03} Paturel, G., et al.\ 2003, \aap, 412, 45
\bibitem[Peterson et al.(2004)]{pet04} Peterson, B.~M., et al.\ 2004, \apj, 613, 682
\bibitem[Peterson et al.(2005)]{pet05} Peterson, B.~M., et al.\ 2005, \apj, 632, 799
\bibitem[Shu et al.(2007)]{shu07} Shu, X.~W., Wang, J.~X., Jiang, P., Fan, L.~L., \& Wang, T.~G.\ 2007, \apj, 657, 167
\bibitem[Schmitt(2001)]{sch01} Schmitt, H.~R.\ 2001, \aj, 122, 2243
\bibitem[Sorrentino et al.(2006)]{sor06} Sorrentino, G., Radovich, M., \& Rifatto, A. 2006, \aap, 451, 809 (SRR06)
\bibitem[Stasi{\'n}ska et al.(2006)]{sta06} Stasi{\'n}ska, G., Cid Fernandes, R., Mateus, A., Sodr{\'e}, L., \& Asari, N.~V.\ 2006, \mnras, 371, 972
\bibitem[Tran(2001)]{tra01} Tran, H. D. 2001, \apj, L554, 19
\bibitem[Tran(2003)]{tra03} Tran, H. D. 2003, \apj, 583, 632
\bibitem[Verdes-Montenegro et al.(2001)]{ver01} Verdes-Montenegro et al. 2001,\aap, 377, 812
\end{thebibliography}
\end{document}